\documentclass{article}
\usepackage{spconf,amsmath,graphicx}

\usepackage{empheq}
\usepackage{units}
\usepackage{pgfplots}
\usepackage{tikzscale}
\usepackage{subcaption}
\usepackage{bbold}
\usepackage[utf8]{inputenc}

\usepackage{tabularx}
\newcolumntype{Y}{>{\centering\arraybackslash}X}
\usepackage{booktabs}

\pgfplotsset{compat=newest}
\usepgfplotslibrary{groupplots}
\usepgfplotslibrary{dateplot}

\usepackage[subtle,charwidths]{savetrees}

\title{DNN-Based Speech Presence Probability Estimation\\for Multi-Frame Single-Microphone Speech Enhancement}
\name{Marvin Tammen, D\"{o}rte Fischer, Bernd T. Meyer, Simon Doclo
	\thanks{This work was funded by the Deutsche Forschungsgemeinschaft (DFG, German Research Foundation) – Project ID 390895286 – EXC 2177/1.}}
\address{Department of Medical Physics and Acoustics\\and Cluster of Excellence Hearing4all\\University of Oldenburg, Germany}
\begin{document}
\ninept
\maketitle
\begin{abstract}
Multi-frame approaches for single-microphone speech enhancement, e.g., the multi-frame minimum-power-distortionless-response (MFMPDR) filter, are able to exploit speech correlations across neighboring time frames.
In contrast to single-frame approaches such as the Wiener gain, it has been shown that multi-frame approaches achieve a substantial noise reduction with hardly any speech distortion, provided that an accurate estimate of the correlation matrices and especially the speech interframe correlation (IFC) vector is available.
Typical estimation procedures of the IFC vector require an estimate of the speech presence probability (SPP) in each time-frequency (TF) bin.
In this paper, we propose to use a bi-directional long short-term memory deep neural network (DNN) to estimate the SPP for each TF bin.
Aiming at achieving a robust performance, the DNN is trained for various noise types and within a large signal-to-noise-ratio range.
Experimental results show that the MFMPDR in combination with the proposed data-driven SPP estimator yields an increased speech quality compared to a state-of-the-art model-based SPP estimator. Furthermore, it is confirmed that exploiting interframe correlations in the MFMPDR is beneficial when compared to the Wiener gain especially in adverse scenarios.
\end{abstract}

\section{Introduction}
In many hands-free speech communication systems such as hearing aids, mobile phones and smart speakers, ambient noise may degrade the speech quality and intelligibility of the recorded microphone signals. 
Hence, several single- and multi-microphone speech enhancement approaches have been proposed~\cite{vary_digital_2006,hendriks_dft-domain_2013,benesty_speech_2011,doclo_multichannel_2015,vincent_audio_2018}.
Typical single-microphone speech enhancement approaches apply a real-valued spectro-temporal gain, e.g., the Wiener gain (WG)~\cite{vary_digital_2006}, to the noisy short-time Fourier transform (STFT) coefficients to obtain an estimate of the clean speech signal. 
A disadvantage of these methods is that stronger noise reduction typically goes hand-in-hand with increased speech distortion. 

In contrast to these single-frame approaches, multi-frame approaches~\cite{huang_multi-frame_2012,schasse_estimation_2014,andersen_robust_2018,fischer_combined_2016,fischer_robust_2018} apply a complex-valued filter to the noisy STFT coefficients and are able to take into account the speech correlation across consecutive time frames.
Similarly to the minimum-variance-distortionless-response (MVDR) beamformer and the minimum-power-distortionless-response beamformer (MPDR) for multi-microphone speech enhancement~\cite{doclo_multichannel_2015, veen_beamforming:_1988}, a multi-frame MPDR (MFMPDR) filter has been proposed for single-microphone speech enhancement~\cite{huang_multi-frame_2012,schasse_estimation_2014,fischer_robust_2018}.
This multi-frame filter requires an estimate of the noisy correlation matrix and the speech interframe correlation (IFC) vector in each time-frequency (TF) bin.
When oracle estimates of these quantities are available, it has been shown in~\cite{huang_multi-frame_2012, fischer_sensitivity_2017} that the MFMPDR filter achieves a good noise reduction and hardly any speech distortion in contrast to the WG.
However, it has also been shown that the speech enhancement performance is very sensitive to estimation errors of the highly time-varying speech IFC vector~\cite{fischer_sensitivity_2017}.

In~\cite{schasse_estimation_2014} a maximum likelihood (ML)-based approach has been proposed to estimate the speech IFC vector from the noisy microphone signals.
The ML estimator typically requires an estimate of the speech presence probability (SPP) in each TF bin.
Several model-based SPP estimators have been proposed~\cite{cohen_noise_2003,gerkmann_improved_2008,souden_gaussian_2010,gerkmann_unbiased_2012} based on the assumption that the speech and noise STFT coefficients are uncorrelated, complex Gaussian distributed random variables.
These estimators, however, have difficulties with accurately estimating the SPP in the short STFT frames that are required to capture the highly time-varying speech IFC vector. 

In recent years, data-driven supervised learning-based approaches have gained a lot of attention in a multitude of applications, including single-microphone speech enhancement~\cite{wang_ideal_2005, wang_training_2014, williamson_complex_2016, kolbaek_multitalker_2017, weninger_speech_2015, chazan_deep_2017}.
A common approach is to estimate real-valued TF masks, which are applied to the noisy STFT coefficients.
Furthermore, mask-based approaches have been recently proposed to estimate the speech and noise correlation matrices that are required by multi-microphone speech enhancement approaches such as the MVDR beamformer or the generalized eigenvalue beamformer~\cite{heymann_neural_2016, jiang_robust_2018}.

Inspired by the approach in~\cite{heymann_neural_2016}, in this paper we propose to use a data-driven SPP to estimate the required speech IFC vector for the MFMPDR filter.
More in particular, we use a bidirectional long short-term memory (BLSTM)~\cite{graves_framewise_2005} deep neural network (DNN) to estimate the SPP in each TF bin given the noisy STFT coefficients.
Aiming at achieving a robust performance, the DNN is trained on the WSJ0~\cite{paul_design_1992} and NOISEX92~\cite{varga_assessment_1993} datasets using a signal-to-noise ratio (SNR) range from 0 to \unit[20]{dB}.
Experimental results for non-matched noise types and partially non-matched SNRs show that using the proposed DNN-based SPP estimate yields a larger speech quality improvement compared to the model-based SPP estimate~\cite{gerkmann_unbiased_2012}. 
Furthermore, when utilizing either of the SPP estimates to implement an MFMPDR or a WG, the benefit of exploiting speech IFCs is confirmed~\cite{schasse_estimation_2014,fischer_combined_2016}.


\section{Signal Model}
\label{sec:signal model}
We consider an acoustic scenario with one speech source and ambient noise, recorded using a single microphone.
In the STFT domain, the noisy microphone signal is given by
\begin{equation}
Y(k,\ l) = X(k,\ l) + N(k,\ l),
\label{eq:model}
\end{equation}
where $X(k,\ l)$ denotes the speech component and $N(k,\ l)$ denotes the noise component at the $k$-th frequency bin and the $l$-th time frame.
Multi-frame speech enhancement approaches~\cite{huang_multi-frame_2012,schasse_estimation_2014,andersen_robust_2018,fischer_robust_2018} estimate the speech component by applying a finite impulse response filter with $N$ taps to the noisy STFT coefficients, i.e.,
\begin{equation}
\widehat{X}(k,\ l) = \sum_{n=0}^{N-1}H^*_n(k,\ l) Y(k,\ l-n),
\label{eq: FIR}
\end{equation} 
where $\widehat{\circ}$ denotes an estimate of $\circ$, $H_n(k,\ l)$ denotes the $n$-th filter coefficient, and $^*$ denotes the complex-conjugate operator.
Using vector notation, \eqref{eq:model} and \eqref{eq: FIR} can be written as
\begin{align}
\mathbf{y}(k,\ l) &= \mathbf{x}(k,\ l) + \mathbf{n}(k,\ l)\label{eq:microphone_signal}\\
\widehat{X}(k,\ l) &= \mathbf{h}^H(k,\ l) \mathbf{y}(k,\ l),
\label{eq: FIR vector}
\end{align}
where $^H$ denotes the Hermitian operator and the $N$-dimensional vectors $\mathbf{h}(k,l)$ and $\mathbf{y}(k,l)$ contain the filter coefficients and $N$ consecutive STFT coefficients, i.e.,
\begin{align}
\mathbf{h}(k,\ l) &= [H_0(k,\ l),\ H_1(k,\ l),\ \dots,\ H_{N-1}(k,\ l)]^T,\\
\mathbf{y}(k,l) &= [Y(k,\ l),\ Y(k,\ l-1),\ \dots,\ Y(k,\ l-N+1)]^T.
\end{align}
This is analogous to multi-microphone beamforming approaches~\cite{doclo_multichannel_2015,vincent_audio_2018,veen_beamforming:_1988} by considering the FIR filter as a spatial filter and frames as microphone inputs.
Since all frequency bins are treated individually, in the remainder of this paper we omit the frequency index $k$.

Assuming that the speech and noise components are uncorrelated, the noisy correlation matrix $\boldsymbol{\Phi_y}(l) = \mathcal{E} \left\lbrace\mathbf{y}(l) \mathbf{y}^H(l)\right\rbrace$, with $\mathcal{E}\{\circ\}$ the expectation operator, can be written as
\begin{equation}
\boldsymbol{\Phi_y}(l) = \boldsymbol{\Phi_x}(l) + \boldsymbol{\Phi_n}(l),
\label{eq:mic correlation matrix}
\end{equation}
with the speech and noise correlation matrices $\boldsymbol{\Phi_x}(l) = \mathcal{E} \left\lbrace\mathbf{x}(l) \mathbf{x}^H(l)\right\rbrace$ and $\boldsymbol{\Phi_n}(l) = \mathcal{E} \left\lbrace\mathbf{n}(l) \mathbf{n}^H(l)\right\rbrace$.
In~\cite{huang_multi-frame_2012}, it has been proposed to exploit the speech correlation across consecutive time frames by separating the speech component into a correlated and an uncorrelated part, i.e.,
\begin{equation}
\mathbf{x}(l) = \underbrace{\boldsymbol{\gamma_x}(l) X(l)}_{\mathclap{\mathrm{correlated}}} \quad + \quad \underbrace{\mathbf{x'}(l)}_{\mathclap{\mathrm{uncorrelated}}},
\label{eq:correlated uncorrelated}
\end{equation}
where the (highly time-varying) normalized speech IFC vector $\boldsymbol{\gamma_x}(l)$ describes the correlation between the current and previous time frames w.r.t. the speech STFT coefficient $X(l)$, i.e.,
\begin{equation}
\boldsymbol{\gamma_x}(l) = \frac{\mathcal{E}\left\lbrace \mathbf{x}(l) X^*(l) \right\rbrace}{\mathcal{E}\left\lbrace \left| X(l) \right|^2 \right\rbrace}
= \frac{\boldsymbol{\Phi}_{\mathbf{x}}(l) \mathbf{e}}{\mathbf{e}^T \boldsymbol{\Phi}_{\mathbf{x}}(l) \mathbf{e}},
\label{eq:gamma_x}
\end{equation}
with the vector $\mathbf{e}$ selecting the first column of $\boldsymbol{\Phi}_{\mathbf{x}}(l)$ and $\mathbf{e}^T \boldsymbol{\Phi}_{\mathbf{x}}(l) \mathbf{e} = \phi_X(l) = \mathcal{E} \left\lbrace \left| X(l) \right|^2 \right\rbrace$ the speech power spectral density (PSD).
Note that since $X(l)$ is fully correlated with itself, the first element of the speech IFC vector $\boldsymbol{\gamma_x}(l)$ in \eqref{eq:gamma_x} is equal to 1, such that the first element of the uncorrelated speech vector $\mathbf{x'}(l)$ is equal to 0.
Substituting \eqref{eq:correlated uncorrelated} in \eqref{eq:microphone_signal}, we obtain the multi-frame signal model
\begin{align}
	\mathbf{y}(l) = \boldsymbol{\gamma_x}(l) X(l) + \mathbf{x'}(l) + \mathbf{n}(l),
	\label{eq:multi_frame_signal_model}
\end{align}
where the uncorrelated speech component $\mathbf{x'}$ is treated as an interference.

Similarly to the speech IFC vector in \eqref{eq:gamma_x}, the noisy IFC vector and the noise IFC vector can be defined as
\begin{equation}
\boldsymbol{\gamma_y}(l) = \frac{\boldsymbol{\Phi}_{\mathbf{y}}(l) \mathbf{e}}{\mathbf{e}^T \boldsymbol{\Phi}_{\mathbf{y}}(l) \mathbf{e}}, \quad \boldsymbol{\gamma_n}(l) = \frac{\boldsymbol{\Phi}_{\mathbf{n}}(l) \mathbf{e}}{\mathbf{e}^T \boldsymbol{\Phi}_{\mathbf{n}}(l) \mathbf{e}},
\label{eq: noisy and noise IFC vectors}
\end{equation}
with $\mathbf{e}^T \boldsymbol{\Phi}_{\mathbf{y}}(l) \mathbf{e} = \mathcal{E} \left\lbrace \left| Y(l) \right|^2 \right\rbrace$ and $\mathbf{e}^T \boldsymbol{\Phi}_{\mathbf{n}}(l) \mathbf{e} = \phi_N(l) = \mathcal{E} \left\lbrace \left| N(l) \right|^2 \right\rbrace$ denoting the noisy and noise PSDs, respectively.
Using \eqref{eq: noisy and noise IFC vectors} in \eqref{eq:mic correlation matrix}, the speech IFC vector $\boldsymbol{\gamma}_\mathbf{x}(l)$ can be obtained as
\begin{empheq}[]{align}
\boldsymbol{\gamma}_\mathbf{x}(l) = \frac{1 + \xi(l)}{\xi(l)} \boldsymbol{\gamma_y}(l) - \frac{1}{\xi(l)} \boldsymbol{\gamma_n}(l),
\label{eq:speech IFC vector}
\end{empheq}
with the a-priori SNR $\xi(l) = \frac{\phi_X(l)}{\phi_N(l)}$.

\section{Multi-Frame MPDR Filter}
\label{sec:MFMPDR}
In \cite{huang_multi-frame_2012}, the MFMPDR filter for single-microphone speech enhancement was proposed, which aims at minimizing the output PSD while preserving the correlated speech component.
The corresponding constrained optimization problem is given by
\begin{empheq}{align}
\min_{\mathbf{h}(l) \, \in \, \mathbb{C}^N} \quad \mathbf{h}^H(l) \boldsymbol{\Phi}_{\mathbf{y}}(l) \mathbf{h}(l), \quad
\text{s.t. } \ \mathbf{h}^H(l) \boldsymbol{\gamma_x}(l) = 1.
\end{empheq}
Solving this problem, the filter vector is equal to~\cite{huang_multi-frame_2012}
\begin{empheq}[box=\fbox]{align}
\mathbf{h}_\text{MFMPDR} = \frac{\boldsymbol{\Phi}^{-1}_{\mathbf{y}}(l) \boldsymbol{\gamma_x}(l)}{\boldsymbol{\gamma}^H_\mathbf{x}(l) \boldsymbol{\Phi}^{-1}_{\mathbf{y}}(l) \boldsymbol{\gamma_x}(l)}
\label{eq:MFMPDR}
\end{empheq}

%

\section{Parameter Estimation}
\label{sec:parameter_estimation}
\begin{figure}[t]
	\centering
	\includegraphics[width=\linewidth]{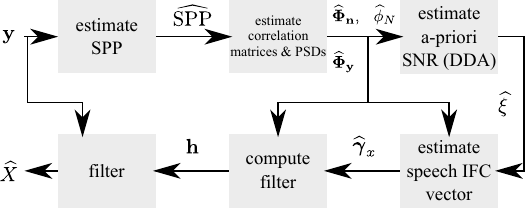}
	\caption{Diagram of parameter estimation and multi-frame filtering.}
	\label{fig:diagram}
\end{figure}

In practice, the performance of the MFMPDR filter depends on how well the time-varying correlation matrix $\boldsymbol{\Phi_y}(l)$ as well as the highly time-varying speech IFC vector $\boldsymbol{\gamma_x}(l)$ can be estimated from the noisy microphone signals.
In \cite{fischer_sensitivity_2017} it has been shown that the performance of the MFMPDR filter is very sensitive to estimation errors of the speech IFC vector.
Whereas estimating the noisy correlation matrix $\boldsymbol{\Phi}_{\mathbf{y}}(l)$ is rather straightforward, accurately estimating the speech IFC vector $\boldsymbol{\gamma_x}(l)$ is not so trivial~\cite{schasse_estimation_2014,andersen_robust_2018,fischer_robust_2018,fischer_sensitivity_2017}.
Typically, this vector requires an estimate of the a-priori SNR $\xi(l)$ and the noise correlation matrix $\boldsymbol{\Phi}_{\mathbf{n}}(l)$, which in turn require an estimate of the SPP in each TF bin~\cite{schasse_estimation_2014}.
The following subsections discuss the estimation of the noisy and noise correlation matrices, the speech IFC vector, as well as the a-priori SNR using either a state-of-the-art model-based SPP estimator or the proposed DNN-based SPP estimator.
Fig.~\ref{fig:diagram} depicts the parameter estimation and multi-frame filtering process.

\subsection{Correlation Matrices Estimation}
\label{sec:correlation matrices}
The noisy correlation matrix $\boldsymbol{\Phi}_{\mathbf{y}}(l)$ is estimated using recursive smoothing with smoothing constant $\lambda_y$, i.e.,
\begin{equation}
\widehat{\boldsymbol{\Phi}}_{\mathbf{y}}(l) = \lambda_y \widehat{\boldsymbol{\Phi}}_{\mathbf{y}}(l-1) + (1-\lambda_y) \mathbf{y}(l) \mathbf{y}^H(l).
\label{eq:noisy correlation}
\end{equation}
To estimate the noise correlation matrix $\boldsymbol{\Phi_n}(l)$, similarly to \cite{souden_integrated_2011} we apply a recursive smoothing procedure to the noisy microphone signals, where the smoothing factor for each TF bin depends on a time-varying SPP estimate $\widehat{\mathrm{SPP}}(l)$ and a smoothing constant $\alpha_n$, i.e.,
\begin{empheq}[]{align}
\label{eq:noise correlation matrix}
\widehat{\boldsymbol{\Phi}}_{\mathbf{n}}(l) &= \lambda_n(l) \widehat{\boldsymbol{\Phi}}_{\mathbf{n}}(l-1) + (1-\lambda_n(l)) \mathbf{y}(l) \mathbf{y}^H(l)\\
\lambda_n(l) &= \alpha_n + (1-\alpha_n) \widehat{\mathrm{SPP}}(l).\label{eq:lambda_n}
\end{empheq}
In the limiting cases, we have
\begin{empheq}[left=\empheqlbrace]{align}
\widehat{\mathrm{SPP}}(l) &= 0 \Rightarrow \lambda_n(l) = \alpha_n\\
\widehat{\mathrm{SPP}}(l) &= 1 \Rightarrow \lambda_n(l) = 1 \Rightarrow \widehat{\boldsymbol{\Phi}}_{\mathbf{n}}(l) = \widehat{\boldsymbol{\Phi}}_{\mathbf{n}}(l-1).
\end{empheq}
We consider two approaches to estimate the SPP for each TF bin required in \eqref{eq:lambda_n}.
As the reference, denoted with subscript $\circ_\mathrm{R}$, we use the model-based approach from \cite{gerkmann_unbiased_2012}, which assumes that the speech and noise STFT coefficients are complex Gaussian distributed.
Using this assumption, likelihood functions for speech presence and speech absence can be derived, yielding the SPP estimate
\begin{empheq}[box=\fbox]{align}
\label{eq:SPP_R}
\widehat{\mathrm{SPP}}_\mathrm{R}(l) = \left( 1 + \frac{P(\mathcal{H}_0)}{P(\mathcal{H}_1)}(1 + \xi_{\mathcal{H}_1}) e^{- \frac{|Y(l)|^2}{\widehat{\phi}_N(l-1)} \frac{\xi_{\mathcal{H}_1}}{1 + \xi_{\mathcal{H}_1}}} \right)^{-1}
\end{empheq}
where $P(\mathcal{H}_1)$ and $P(\mathcal{H}_0)$ denote the prior probability of speech presence and absence, respectively, and the parameter 
$\xi_{\mathcal{H}_1}$ denotes a typical a-priori SNR encountered during speech presence.
Note that this method relies on the noise PSD estimate of the \emph{previous} frame $\widehat{\phi}_N(l-1) = \mathbf{e}^T \widehat{\boldsymbol{\Phi}}_{\mathbf{n}}(l-1)\mathbf{e}$.

Alternatively, in this paper we propose to exploit the capabilities of a BLSTM DNN to capture temporal and spectral structures in order to estimate the SPP. 
The DNN is trained to perform a mapping between the noisy STFT coefficient magnitudes and the SPP, i.e.,
\begin{empheq}[box=\fbox]{align}
	\widehat{\mathrm{SPP}}_\mathrm{DNN}(l) = f_{\boldsymbol{\Theta}} \left\lbrace \mathbf{\left\lvert Y \right\rvert} \right\rbrace (l)
	\label{eq:SPP_DNN_estimate}
\end{empheq}
with $\mathbf{\left\lvert Y \right\rvert} \in \mathbb{R}^{K \times L}$ containing all $K$ frequency bins and $L$ time frames of the noisy STFT coefficient magnitudes of the considered signal, $f_{\boldsymbol{\Theta}}$ the trained DNN with parameters $\boldsymbol{\Theta}$, and $\widehat{\mathrm{SPP}}_\mathrm{DNN}(l)$ the DNN-based SPP estimate.
The training process is detailed in Sec.~\ref{sec:training_process}.

\subsection{Speech IFC Vector Estimation}
\label{sec:speech IFC vector]}
Similarly to \eqref{eq:speech IFC vector}, the ML-based approach in \cite{schasse_estimation_2014} estimates the speech IFC vector as
\begin{empheq}[box=\fbox]{align}
\label{eq:ML speech IFC estimate}
\widehat{\boldsymbol{\gamma}}^{\boldsymbol{\mu}}_\mathbf{x}(l) = \frac{1 + \widehat{\xi}(l)}{\widehat{\xi}(l)} \widehat{\boldsymbol{\gamma}}_\mathbf{y}(l) - \frac{1}{\widehat{\xi}(l)} \boldsymbol{\mu_{\gamma_n}}
\end{empheq}
where $\widehat{\xi}(l)$ is an estimate of the a-priori SNR and $\widehat{\boldsymbol{\gamma}}_\mathbf{y}(l)$ is an estimate of the noisy IFC vector obtained similarly as in \eqref{eq: noisy and noise IFC vectors} using $\widehat{\boldsymbol{\Phi}}_{\mathbf{y}}(l)$ from \eqref{eq:noisy correlation}.
The fixed mean noise IFC vector $\boldsymbol{\mu_{\gamma_n}}$ can be computed based on the analysis window and overlap settings~\cite{schasse_estimation_2014}.

Alternatively, by replacing the fixed mean noise IFC vector by a TF-varying noise IFC vector estimate $\widehat{\boldsymbol{\gamma}}_\mathbf{n}(l)$, the speech IFC vector can be computed as
\begin{empheq}[box=\fbox]{align}
	\label{eq:gammax gamman}
	\widehat{\boldsymbol{\gamma}}^{\boldsymbol{\gamma}}_\mathbf{x}(l) = \frac{1 + \widehat{\xi}(l)}{\widehat{\xi}(l)} \widehat{\boldsymbol{\gamma}}_\mathbf{y}(l) - \frac{1}{\widehat{\xi}(l)} \widehat{\boldsymbol{\gamma}}_\mathbf{n}(l)
\end{empheq}
where $\widehat{\boldsymbol{\gamma}}_\mathbf{n}(l)$ is obtained similarly to \eqref{eq: noisy and noise IFC vectors} using $\widehat{\boldsymbol{\Phi}}_{\mathbf{n}}(l)$ from \eqref{eq:noise correlation matrix}.

To estimate the a-priori SNR $\xi(l)$, we apply the well-known decision-directed approach (DDA)~\cite{ephraim_speech_1984}, i.e.,
\begin{equation}
\widehat{\xi}(l) = \lambda_{\mathrm{DDA}} \frac{\widehat{X}(l-1)}{\widehat{\phi}_N(l-1)} + (1-\lambda_{\mathrm{DDA}}) \frac{|Y(l)|^2}{\widehat{\phi}_N(l-1)},
\label{eq:DDA}
\end{equation}
with weighting constant $\lambda_{\mathrm{DDA}}$ and $\widehat{X}(l-1)$ denoting the speech estimate of the previous frame.\pagebreak

\section{DNN Training Process}
\label{sec:training_process}
As described in~\eqref{eq:SPP_DNN_estimate}, the DNN is trained to map the input features, i.e., the noisy STFT coefficient magnitudes, to the SPP. 
More specifically, we train the DNN with the target defined as
\begin{empheq}{align}
\label{eq:SPP_DNN}
\mathrm{SPP}_\mathrm{DNN}(l) = \left( 1 + \frac{P(\mathcal{H}_0)}{P(\mathcal{H}_1)}(1 + \xi_{\mathcal{H}_1}) e^{- \frac{|Y(l)|^2}{\phi_N(l)} \frac{\xi_{\mathcal{H}_1}}{1 + \xi_{\mathcal{H}_1}}} \right)^{-1}.
\end{empheq}
For this target, we compute the noise PSD $\phi_{N}(l)$ via recursive averaging of the noise component, which is available during training, i.e.,
\begin{equation}
\phi_{N}(l) = \alpha_n \phi_{N}(l-1) + (1-\alpha_n) \left\lvert N(l) \right\rvert^2.
\label{eq:phin_DNN}
\end{equation}
As loss function, we use the mean-squared difference between the target SPP defined in~\eqref{eq:SPP_DNN} and the estimated SPP $\widetilde{\mathrm{SPP}}_\mathrm{DNN}(k,\ l)$, i.e.,
\begin{equation}
\label{eq:loss}
\hspace*{-5pt}\frac{1}{L K} \sum_{l=0}^{L-1} \sum_{k=0}^{K-1} \left( \widetilde{\mathrm{SPP}}_\mathrm{DNN}(k,\ l) - \mathrm{SPP}_\mathrm{DNN}(k,\ l) \right)^2,
\end{equation}
where $\widetilde{\mathrm{SPP}}_\mathrm{DNN}(k,\ l) = f_{\boldsymbol{\tilde{\Theta}}} \left\lbrace \mathbf{\left\lvert Y \right\rvert} \right\rbrace (k,\ l)$ uses the current set of parameters $\boldsymbol{\tilde{\Theta}}$.
The DNN is composed of an input layer with 33 input nodes, a hidden BLSTM layer with 256 nodes for each direction, two hidden fully-connected layers with 513 nodes each, and an output layer with 33 nodes.
The corresponding activation functions of the hidden and output layers are $\tanh$, rectifying linear unit ($\mathrm{ReLU}$), $\mathrm{ReLU}$, and $\mathrm{sigmoid}$, respectively, inherently restricting the SPP estimates to $]0,\ 1[$.
This network architecture is inspired by the DNN used in \cite{heymann_neural_2016} and has been tested for various sets of hyperparameters.

The network weights are initialized using a uniform distribution $U(-a,\ a)$, with $a = \sqrt{6 / (n_\text{in} + n_\text{out})}$, and $n_\text{in}$ and $n_\text{out}$ the number of input and output neurons of the layer, respectively~\cite{glorot_understanding_2010}.
All bias values are initialized with 0.
To decrease the dynamic range of the input data and to stabilize the training process, we apply batch normalization to the input and before the activations of the hidden layers~\cite{ioffe_batch_2015}.
To optimize the network parameters, the Adam optimizer is utilized with parameters as proposed in~\cite{kingma_adam:_2014}, with the learning rate set to $10^{-3}$ and the smoothing parameters for the gradient and the squared gradient set to 0.9 and 0.999, respectively.
If the $l^2$-norm of a gradient is larger than 1, the gradient is divided by this norm.

To evaluate the model performance, we make use of a separate validation set as described in Sec.~\ref{sec:dataset}.
The training is stopped either after 100 epochs or after the validation loss as measured by~\eqref{eq:loss} has not decreased for 5 epochs.
The DNN is implemented in PyTorch 1.2.0~\cite{paszke_automatic_2017}, and training and evaluation are performed on a multi-GPU system utilizing 3 NVIDIA GeForce\textsuperscript{\textregistered} GTX 1080 Ti graphics cards.

\section{Experimental Results}
\label{sec:experimental_results}
In this section, we compare the speech enhancement performance of the MFMPDR filter in \eqref{eq:MFMPDR} using 
\begin{enumerate}
	\item to estimate the SPP required in~\eqref{eq:lambda_n}: either the model-based SPP estimator $\widehat{\mathrm{SPP}}_\mathrm{R}$ in \eqref{eq:SPP_R} or the proposed DNN-based estimator $\widehat{\mathrm{SPP}}_\mathrm{DNN}$ in \eqref{eq:SPP_DNN_estimate}.
	\item to estimate the speech IFC vector: either the fixed mean noise IFC vector $\boldsymbol{\mu_{\gamma_n}}$ in \eqref{eq:ML speech IFC estimate} or the estimated time-varying noise IFC vector $\widehat{\boldsymbol{\gamma}}_\mathbf{n}(l)$ in~\eqref{eq:gammax gamman}.
\end{enumerate}
In addition, to investigate the impact of exploiting speech IFCs, we also use the SPP estimators in~\eqref{eq:SPP_R} and~\eqref{eq:SPP_DNN_estimate} in a (single-frame) Wiener gain (WG), resulting in a total of 6 compared methods.

\subsection{Dataset}
\label{sec:dataset}
As clean speech material, we have used the training, development, and test sets of the WSJ0 corpus~\cite{paul_design_1992} for training, model validation, and evaluation, respectively.
The noisy microphone signals have been generated by adding scaled (randomly chosen) noise segments to the clean speech signals at a sampling frequency of \unit[16]{kHz}.
Regarding noise, we have used the NOISEX92 database~\cite{varga_assessment_1993} for training and the Aurora database~\cite{hirsch_aurora_2000} for evaluation, resulting in a strong mismatch between training and evaluation conditions in order to evaluate the generalization capability of the proposed method.
For each training utterance, the corresponding broadband SNR has been uniformly sampled from $\unit[{[0, 20]}]{dB}$.
For evaluation, 4 random utterances from the WSJ0 test set have been used at broadband \break SNRs $ \in \unit[\{-5,\ 0,\ 5,\ 10,\ 15,\ 20\}]{dB}$ for each of the 8 noise types in the Aurora database~\cite{hirsch_aurora_2000}.
In total, this results in 12776, 2348, and 192 utterances for training, validation, and evaluation, respectively.

\subsection{Simulation Settings}
Since the speech IFC vector is highly time-varying, we employ an STFT with a high temporal resolution, i.e., a frame length of \unit[4]{ms} and a frame shift of \unit[1]{ms}, similarly as in \cite{huang_multi-frame_2012, schasse_estimation_2014, fischer_combined_2016,fischer_robust_2018}.
A Hann window is used for both STFT analysis and synthesis.
The parameters of both the model-based SPP estimator $\widehat{\mathrm{SPP}}_\mathrm{R}$ in \eqref{eq:SPP_R} and the DNN-based SPP estimator $\widehat{\mathrm{SPP}}_\mathrm{DNN}$ in \eqref{eq:SPP_DNN_estimate} are set as proposed in~\cite{gerkmann_unbiased_2012}, i.e., $P(\mathcal{H}_1) = P(\mathcal{H}_0) = 0.5$ and $\xi_{\mathcal{H}_1} = \unit[15]{dB}$.
As recursive smoothing constants, we use $\alpha_n = 0.98$, $\lambda_y = 0.92$, and $\lambda_{\mathrm{DDA}} = 0.97$.
The MFMPDR filters use a filter length of $N = 18$, such that correlations within a window of \unit[21]{ms} can be exploited.
To be more comparable to the MFMPDR filters, the WG methods are used with the same settings, except for $N=1$.
To improve numerical stability when inverting a matrix, we perform regularization using diagonal loading as in \cite{huang_multi-frame_2012, schasse_estimation_2014} with regularization parameter $\delta = 10^{-3}$.
Finally, all compared methods use a minimum gain of \unit[-17]{dB}.

\subsection{Results}
\begin{table}
	\begin{tabularx}{\linewidth}{l||c|c|c|c|c|c}
		\toprule[2pt]
		SNR / dB & -5 & 0 & 5 & 10 & 15 & 20\\ 
		\midrule[2pt]
		$\textrm{MFMPDR}_{\mathrm{R},\mu}$ & 0.09 & 0.27 & 0.31 & 0.31 & 0.27 & 0.24 \\ 
		\hline 
		$\textrm{MFMPDR}_{\mathrm{DNN},\mu}$ & \textbf{0.22} & \textbf{0.33} & \textbf{0.41} & \textbf{0.35} & \textbf{0.33} & 0.22 \\ 
		\hline
		$\textrm{MFMPDR}_{\mathrm{R},\gamma}$ & -0.01 & 0.15 & 0.24 & 0.29 & 0.29 & 0.24 \\
		\hline
		$\textrm{MFMPDR}_{\mathrm{DNN},\gamma}$ & 0.04 & 0.21 & 0.29 & 0.32 & 0.29 & 0.21 \\ 
		\hline
		$\textrm{WG}_{\mathrm{R}}$ & 0.04 & 0.13 & 0.18 & 0.23 & 0.26 & \textbf{0.28} \\
		\hline
		$\textrm{WG}_{\mathrm{DNN}}$ & 0.06 & 0.16 & 0.20 & 0.24 & 0.27 & 0.23 \\
	\end{tabularx}
	\caption{PESQ / MOS improvements vs. input SNR / dB, averaged over all evaluation set utterances and noise types.}
	\label{tab:pesq_results}
\end{table}
For the 6 considered methods, Tab.~\ref{tab:pesq_results} depicts the improvements in terms of the perceptual evaluation of speech quality (PESQ)~\cite{itu-t_perceptual_2001} measure w.r.t. the noisy microphone signals as a function of the input SNR.
The clean speech signal has been used as the reference signal.
Subscripts denote which SPP estimator was used and, in the case of the MFMPDR filters, whether the mean IFC noise vector $\boldsymbol{\mu_{\gamma_n}}$ or the time-varying noise IFC vector $\boldsymbol{\gamma_n}(l)$ was utilized.
The presented values are averaged over all utterances and noise types included in the evaluation set.

First, it can be observed that the MFMPDR filter utilizing the proposed DNN-based SPP estimate $\widehat{\mathrm{SPP}}_\mathrm{DNN}(l)$ and the fixed mean noise IFC vector $\boldsymbol{\mu_{\gamma_n}}$ ($\textrm{MFMPDR}_{\mathrm{DNN},\mu}$), yields the highest PESQ improvements for all input SNRs except \unit[20]{dB}. 
Second, comparing the methods utilizing either the model-based SPP estimate $\widehat{\mathrm{SPP}}_\mathrm{R}(l)$ or the DNN-based SPP estimate $\widehat{\mathrm{SPP}}_\mathrm{DNN}(l)$, the advantages of using the DNN-based estimator are evident. This may be explained by the fact that, in contrast to the model-based estimator, the DNN can exploit spectral structures of speech and noise.
Third, contrasting the MFMPDR filters and the WG, it can be confirmed that exploiting speech IFCs may yield higher speech quality improvements than directly using the SPP estimate in a WG approach~\cite{schasse_estimation_2014,fischer_combined_2016} (except for an input SNR of \unit[20]{dB}). The difference between the MFMPDR-based methods and the WG-based methods increases for lower input SNRs, suggesting that exploiting the speech IFCs is especially helpful in adverse scenarios. 
Fourth, using the fixed mean noise IFC vector in the MFMPDR filters consistently leads to larger PESQ improvements than using the estimated time-varying noise IFC vector. 
Considering the results in \cite{schasse_estimation_2014,fischer_comparison_2019}, this suggests that a filter bank with higher frequency resolution is required to effectively incorporate an estimate of the time-varying noise IFC vector into the MFMPDR filter.

\section{Conclusion}
\label{sec:conclusion}
In this paper we considered a DNN-based SPP estimator for multi-frame approaches in single-microphone speech enhancement.
Since the MFMPDR filter requires accurate estimates of the time-varying noisy correlation matrix and especially the speech IFC vector, in this paper we propose to use a DNN to improve the estimation of the speech IFC vector.
The DNN is trained to map noisy STFT coefficient magnitudes to an SPP on a database comprising multiple noise types within a large SNR range to improve the generalization capability of the DNN.
We demonstrate a higher objective speech quality improvement when using the proposed DNN-based SPP estimator instead of a state-of-the-art model-based estimator.
Furthermore, by comparing the MFMPDR filters with Wiener gains based on equal SPP estimates, we confirm that utilizing interframe correlations can be beneficial especially in adverse scenarios.
\bibliographystyle{IEEEbib}
{\bibliography{strings,My_Library}}

\end{document}